# A New Class of MDS Erasure Codes Based on Graphs

Nattakan Puttarak†, Phisan Kaewprapha†, Ng Boon Chong‡ and Jing Li (Tiffany)†
† Department of Electrical and Computer Engineering, Lehigh University, Bethlehem, PA 18015
‡ College of Electrical and Electronic Engineering, Nanyang Technological University, Singapore 138664
Emails: {nap205, phk205, jingli}@lehigh.edu, EBCNg@ntu.edu.sg

*Abstract*— Maximum distance separable (MDS) array codes are XOR-based optimal erasure codes that are particularly suitable for use in disk arrays. This paper develops an innovative method to build MDS array codes from an elegant class of nested graphs, termed *complete-graph-of-rings (CGR)*. We discuss a systematic and concrete way to transfer these graphs to array codes, unveil an interesting relation between the proposed map and the renowned perfect 1-factorization, and show that the proposed CGR codes subsume B-codes as their "contracted" codes. These new codes, termed *CGR codes*, and their dual codes are simple to describe, and require minimal encoding and decoding complexity.

## I. INTRODUCTION

Now a widely-adopted industrial standard *redundant/reliable array of inexpensive/independent disks* (RAID) allows data to be stored in a redundant, distributive, and balanced way to improve the overall performance, including fault tolerance, input/output (I/O) throughput and scalability. Central to RAID is the availability of efficient erasure codes. An $(N, K, d_{min})$ erasure code with $t$-erasure correcting capability is considered (spatially) optimal, if it is maximum distance separable (MDS) or achieves the singleton bounds: $t = d_{min} - 1 = N - K$. Reed Solomon (RS) codes, the renowned class of optimal erasure codes, are unfortunately un-fitful for RAID, due to their expensive Galois field operations and the dense Tanner-graph structure. The latter is particularly costly in RAID, because to encode a parity symbol (generally mapped to a spare disk) requires the reading of *many* data symbols (data disks), and to decode a data symbol also requires the reading of *many* other data symbols, causing a large I/O overhead [2].

MDS codes are difficult to construct. Many attempts have been made to improve the code efficiency while maintaining a low encoding/decoding complexity [1]- [10]. The invention of EVENODD codes in the last decade [3], [4] is true cause for excitement, because they opened a refreshing way of constructing MDS codes through an elegant *array* layout and simple XOR operations. A sequence of array constructions capable of correcting two, three or four erasures have since been proposed, including the X-codes, the HoVer codes, and the Star-codes [6], [7]. These MDS or near-MDS codes follow the same line of approach as EVENODD codes, namely, parity bits are computed horizontally, vertically, diagonally, or along some lines of some edges in the data array. Another noteworthy class of MDS codes, namely, B-Codes and their dual, demonstrate a different construction philosophy. Proposed in the nineties and revisited by Xu *et al* [1], these codes are shown to find close relation to graphs and the perfect 1-factorization (P1F). Several researchers have also looked into representing arrays codes in sparse parity check matrices, and leveraging low-density parity-check (LDPC) and particularly quasi-cyclic LDPC constructions to build MDS array codes [2].

This paper focuses on building MDS array codes from graphs. Much of the motivation comes from the inspiring notion of "code-on-graph" and the intriguing relations between bipartite Tanner graphs and LDPC codes and between complete graphs and B-codes. However, the graph-code relation we explore here is new and different. We will demonstrate a new class of graphs, constructed by embedding *ring graphs* in *complete graphs* and referred to as *complete-graph-of-rings* (CGR). We will provide a systematic way of mapping CGRs to array codes which achieve the singleton bounds and which are capable of correcting up to $2n + 1$ erasures for $n = 1, 2, 3, ....$. The CGR codes we proposed here have extended from our work in [8], which projected the existence of CGR-MDS codes and demonstrated a rich set of code examples through constraint computer search. Here we develop a *concrete* method to construct MDS array codes from CGR graphs. We show that our codes can be concisely and completely described by a simple vector, termed the *offset vector*. We further show that the offset vector can be derived from a deterministic set of rules and the P1F technique applied to the base graph of CGRs.

We discuss in detail of the construction of these codes and their dual codes, which are also MDS. We also unveil a fascinating relation between our code and a specific class of B-codes, $B_{2n+1}$. We show that $B_{2n+1}$ codes can actually be obtained by stripping off certain rows and columns from our arrays!

## II. COMPLETE-GRAPH-OF-RING CODES

To construct the proposed CGR codes takes three steps: (1) constructing an appropriate CGR graph (Subsection II-A, Algorithm 1), (2) mapping the CGR graph to an array

Supported by National Science Foundation under Grant No. CCF-0635199, by the Commonwealth of Pennsylvania through PITA, and by a research grant from Thales Communications.

code (Subsection II-B, Algorithm 2), and (3) determining the offset vector that cyclically shifts rows in the array code to achieve MDS (Subsection II-C, Algorithm 3). Below we detail the 3-step procedure along with examples.

*A. Constructing CGR Graphs*

We first provide the notations and definitions commonly used in the graph theory, to facilitate our discussion.

**Definition 1:** A **complete graph** is a graph in which each pair of graph vertices is connected by an edge. The complete graph $K_v$ has $v$ vertices and $\frac{v(v-1)}{2}$ edges, with each vertex having a degree of $v-1$.

**Definition 2:** A **ring graph**, or, a **cycle graph**, is a *closed graph* in which each node has degree 2 and is connected only to its neighbors. A ring graph $C_n$ has $n$ vertices and $n$ edges.

**Definition 3:** A **perfect one-factorization** of a graph is a partitioning of its edges into subsets, called factors, such that each factor is a graph of degree one and the union of any two factors forms a Hamiltonian cycle [10].

The proposed CGR graphs, denoted as $CGR(K_{v_1}, C_{v_2})$, is constructed by expanding a complete graph $K_{v_1}$ to a nested graph, by replacing each vertex in $K_{v_1}$ with a ring graph $C_{v_2}$, and replacing each edge connecting two vertices in $K_{v_1}$ with a set of $v_2$ parallel edges connecting the respective vertices in two rings. Fig. 1 illustrates two examples of CGR constructions. These graphs can be viewed as a complete graph whose "super vertices" are rings. Alternatively, they can be regarded as "multi-layer ring graphs" such any two layers of rings are incident by a set of parallel edges.

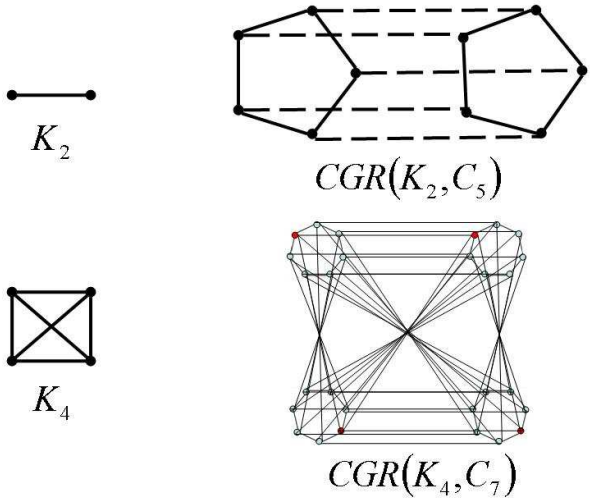

Fig. 1. CGR graphs constructed from base graphs. Left: base graphs $K_2$ and $K_4$; right: resultant CGR graphs $CGR(K_2, C_5)$ and $CGR(K_4, C_7)$.

Before proceeding further, we first provide a sufficient condition that allows a CGR graph to be potentially convertible to an MDS array code.

**Theorem 1:** If a CGR graph $\Upsilon_{v_1, v_2}$ constructed from a complete graph $K_{v_1}$ and a ring graph $C_{v_2}$ satisfying the following conditions: (1) $v_1$ is even, and (2) $v_2 = v_1 + 3$, then there exists a way to place all the vertices and edges in an array of $\frac{v_2 v_1}{2} \times v_2$. When the vertices are interpreted as data bits and the edges connecting two vertices are interpreted as parities associated with two data bits, the resultant array defines an array code of parameters $(N, K, d_{min}) = (v_1, 2, d_{v_1} - 1)$ capable of correcting up to $(v_2 - 2)$ erasures. Its dual code is a $(v_2, v_2 - 2, 3)$ MDS code capable of correcting up to 2 erasures.

**Algorithm 1: Graph Construction and Labeling** Consider constructing a $(v_1 + 1)$-regular CGR $\Upsilon_{v_1, v_2}$ from a complete graph $K_{v_1}$ and a set of $v_1$ rings $C_{v_2}$, where $v_2$ is even and $v_2 = v_1 + 3$.

1) Take a set of $v_1$ number of rings $C_{v_2}$. Label the vertices of the first ring counterclockwise as $0, 1, \cdots, v_2 - 1$; label the vertices of the next ring similarly as $v_2, v_2+1, \cdots, 2v_2-1$, and so on, until all the rings are labeled. We have altogether $v_1$ rings or $v_1$ sets of vertices, where the vertices of the $j$th ring are labeled by $\mathbf{V}_j = \{jv_2, jv_2+1, \cdots, (j+1)v_2-1\}$, for $j = 0, 1, ..., v_1 - 1$.

2) Each edge inside a ring, termed a *ring* edge, is marked by the pair of vertices on both ends. We have altogether $v_1$ sets of ring edges, where the edges of the $j$th ring are labeled by $\mathbf{E}_j = \{(jv_2, jv_2+1), (jv_2+1, jv_2+2), ..., ((j+1)v_2-2, (j+1)v_2-1), ((j+1)v_2-1, jv_2)\}$, for $j = 0, 1, ...v_1 - 1$.

3) For any pair of rings, connect their indexes using $v_2$ parallel *inter-ring* edges, such that the lowest index of one ring is connected to the lowest index of the other, the next lowest is connected to the next lowest, and so on. We have altogether $v_1(v_1 - 1)/2$ sets of inter-ring edges, labeled respectively as $\mathbf{E}_{i,j} = \{(iv_2, jv_2), (iv_2+1, jv_2+1), ..., ((i+1)v_2-1, (j+1)v_2-1)\}$, for $0 \leq i < j \leq v_1 - 1$.

**Example 1:** An example of labeling the vertices for $CGR(K_2, C_5)$ is shown in Fig. 2. Each vertex has 2 ring edges and $(v_1 - 1)$ inter-ring edges connecting between rings. This graph possesses many desirable properties, including symmetry and regularity (all vertices have the same number of degree $v_1 + 1$).

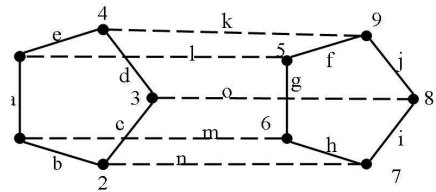

Fig. 2. Labeling of 3-regular $CGR(K_2, C_5)$.

*B. Mapping CGR Graphs to Arrays*

We consider a new graph-code relation, which is different from the usual connection between a Tanner graph and a parity check matrix. Our relation maps the vertices to information bits and edges to parity bits, which are computed by XORing the two information bits on both ends. The resultant code consumes the least possible encoding/decoding complexity (not considering trivial replication), since any parity bit is a function of only two information bits.

The CGR graph constructed in the previous section has altogether $v_1 v_2$ vertices, $v_1 v_2$ ring edges, and $v_1 v_2 (v_1 - 1)/2$ inter-ring edges, all of which can be arranged into an array of size $\frac{v_1(v_1+3)}{2} \times v_2$.

**Algorithm 2: Constructing CGR Array Codes**

Consider constructing a CGR array code using $\text{CGR}(K_{v_1}, C_{v_2})$ labeled by Algorithm 1. The array code will consist of $v_1(v_1+3)/2 = v_1 v_2/2$ rows and $v_2$ columns (recall $v_2 = v_1 + 3$ and $v_1$ even).

1) The $v_1$ sets of vertices, each corresponding to a ring, are placed in the first $v_1$ rows as systematic bits. By default, the vertices in each set is placed in ascending order from left to right to form a row.
2) The $v_1$ sets of ring edges, each corresponding to a ring, are placed in the next $v_1$ rows as parity bits. By default, the edges of the same ring are placed in ascending order, with the one connecting the two smallest indexes being the first, and the wrap-around edge that connects the biggest index and the smallest index being the last.
3) The $v_1(v_1 - 1)/2$ sets of inter-ring edges, each connecting a pair of rings, are placed in the remainder $v_1(v_1-1)/2$ rows as parity bits. The edges in each set is placed in ascending order, with the one connecting the two smallest indexes being the first, and the one connecting the largest indexes being the last.
4) Next, cyclically shift the elements in each row according to an *offset vector*. An offset vector is a predetermined vector in the form of

$$(\alpha_0, \alpha_1, ..., \alpha_{v_1 v_2/2-1}) \in \{0, 1, ..., v_2 - 1\}^{v_1 v_2/2}.$$

Cyclically shift the $j$th row *to the left* by $\alpha_j$ positions, or, equivalent, strip off the first $\alpha_j$ elements in the $j$th row and append them to the end of row. When the offset vector is appropriately designed, such as using Algorithm 3, then the array code is MDS.

**Example 2:** Consider $\text{CGR}(K_2, C_5)$ with vertices labeled from 0 to 9 as shown in Fig. 2. According to Algorithm 2, we can place all the vertices (information bits) and edges (parity of two information bits) in a $5 \times 5$ array, with the first 2 rows for vertices, the next 2 rows for ring edges and the last one row for inter-ring edges. Suppose we are given an offset vector $(0, 1, 2, 2, 4)$, then these five rows should be cyclically shifted by $0,1,2,2,4$ positions to the left, respectively, giving rise to the following array:

| 0 | 1 | 2 | 3 | 4 |
|---|---|---|---|---|
| 6 | 7 | 8 | 9 | **5** |
| $2 \oplus 3$ | $3 \oplus 4$ | $4 \oplus 0$ | $\mathbf{0 \oplus 1}$ | $1 \oplus 2$ |
| $7 \oplus 8$ | $8 \oplus 9$ | $9 \oplus 5$ | $\mathbf{5 \oplus 6}$ | $6 \oplus 7$ |
| $4 \oplus 9$ | $\mathbf{0 \oplus 5}$ | $1 \oplus 6$ | $2 \oplus 7$ | $3 \oplus 8$ |

Each column in the array is interpreted as a symbol, and an erasure occurs to the entire symbol (column). If we physically map each symbol to, for example, a hard disk in RAID, then this array code has stored 2 disks of data in an array of 5 disks, with 3 spare disks serving as redundant protection. This system is robust against up to 3 failed disks, since any two surviving disks suffice to deduce all the information data (easily verifiable). Although information and parity bits are symmetrically distributed across all the symbols, and hence no symbol is purely systematic or parity, this array code is nevertheless regarded as a $(5, 2)$ MDS code with 3-erasure correcting capability.

It is clear from the above example that a judiciously-selected offset vector is a key to ensure MDS. From computer search, one can find many valid offset vectors for any given array. Below we discuss a systematic construction based on P1F, which is guaranteed to generate MDS array codes.

*C. Constructing Offset Vectors Using P1F*

The proposed construction for the offset vector consists of three rules, applicable respectively to the three different types of rows in the array, namely, rows of vertices, rows of ring edges, and rows of inter-ring edges; see Fig. 4.

- <u>Rule 1:</u> The vertex-row corresponding to the $j$th ring, for $j = 0, 1, 2, v_1 - 1$, takes an offset $j$.
- <u>Rule 2:</u> All the ring-edge-rows take the same offset $v_1$.
- <u>Rule 3:</u> The offsets for the rows corresponding to inter-ring edges may be either determined through a computer search or constructed according to Algorithm 3, a P1F-based approach.

We now know that the offset vector for $\text{CGR}(K_{v_1}, C_{v_2})$ may take the form of

$$(\underbrace{0, 1, ..., v_1-1}_{v_1}, \underbrace{v_1, v_1, \cdots, v_1}_{v_1}, \underbrace{\alpha_{2v_1-1}, \alpha_{2v_1}, \cdots, \alpha_{v_1 v_2/2-1}}_{v_1(v_1-1)/2})$$

**Algorithm 3: Determining Offset Vector**

The algorithm determines the offsets for the rows of the inter-ring edges of $\text{CGR}(K_{v_1}, C_{v_2})$, by applying P1F on a larger complete graph $K_{v_1+2}$, and then trimming it down to $K_{v_1}$.

1) First label the vertices in $K_{v_1+2}$ with $0, 1, ..., v_1 - 1$ and $-\infty$ and $+\infty$, where $v_1$ is even.
2) Place all the vertices in a wheel, with $-\infty$ in the center, and all the others in a ring (spaced evenly) surrounding the center. Connect any pair of vertices with an edge.
3) Apply the well-known P1F technique discussed in [10] to group all the edges of $K_{v_1+2}$ in $v_1+1$ factors, such that each factor consists of a center-pointing edge (i.e. edge $(-\infty, i)$ where $i \in \{0, 1, ..., v_1 - 1, \infty\}$) and a set of $v_1/2$ edges that are diagonal to ("perpendicular to") it.

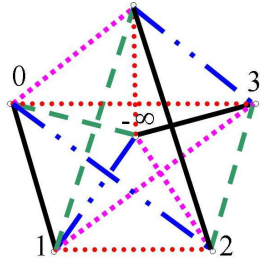

Fig. 3. Complete graph $K_6$.

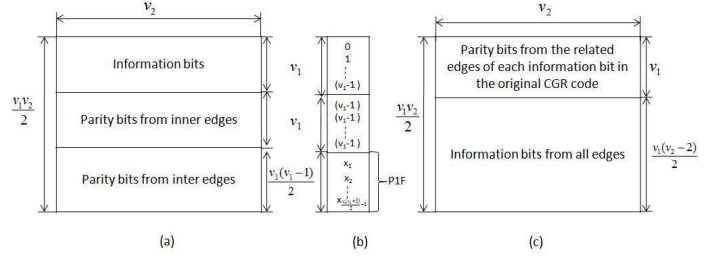

Fig. 4. Structure of CGR code, and its dual code compared to a shift index.

4) Assign the $(-\infty, \infty)$ group an offset $v_1 + 2$, and assign to the other groups distinct offsets chosen arbitrarily from $0, 1, ..., v_1 - 1$.
5) Remove from each factor the edges that are incident with vertices $-\infty$ or $\infty$. What remains are all the edges from the base graph $K_{v_1}$ and their corresponding offsets, which are the offsets for all the inter-ring-edge-rows (Rule 3).

**Example 3:** Consider CGR($K_4, C_7$). To determine the offsets for the inter-ring-edge-rows, consider P1F on $K_6$, as illustrated in Fig. 3. The vertice $-\infty$ is placed in the center, and the vertices $0, 1, 2, 3, +\infty$ may take arbitrary positions in the cycle. The P1F partitions all the edges in 5 factors as shown below.

| center-point edge | diagonal edges | offsetA | offsetB |
|---|---|---|---|
| $(-\infty, +\infty)$ | $(0,3), (1,2)$ | 6 | 6 |
| $(-\infty, 0)$ | $(1, +\infty), (2,3)$ | 1 | 0 |
| $(-\infty, 1)$ | $(0,2), (3, +\infty)$ | 3 | 1 |
| $(-\infty, 2)$ | $(1,3), (0, +\infty)$ | 0 | 2 |
| $(-\infty, 3)$ | $(0,1), (2, +\infty)$ | 2 | 3 |

We have shown two possible assignments of offsets, but there are as many as $v_1(v_1-1)/2$, provided that the $(-\infty, \infty)$ group is always given offset 6. Next, ignore all the edges incident to $-\infty$ or $\infty$. The remainder are the offsets for $(i, j)$, for $0 \le i < j \le v_1 - 1$, which are in fact the offsets for the inter-ring edges connecting the $i$th ring and the $j$th ring. If we place the inter-ring-edge-rows in a natural order in the array, such as

|  | (0,1) | (0,2) | (0,3) | (1,2) | (1,3) | (2,3) |
|---|---|---|---|---|---|---|
| offset1 | 2 | 3 | 6 | 6 | 0 | 1 |
| offset2 | 3 | 1 | 6 | 6 | 2 | 0 |

Now, combining Rule 1 and Rule 2, we have two valid offset vectors for CGR($K_4, K_7$):

offset vector A :$(0, 1, 2, 3, 4, 4, 4, 4, 2, 3, 6, 6, 0, 1)$ (1)

offset vector B :$(0, 1, 2, 3, 4, 4, 4, 4, 3, 1, 6, 6, 2, 0)$ (2)

There are many more, which we cannot possibly list.

**Theorem 2:** Algorithm 1, 2, and 3 combined together construct an MDS array code.

*Proof and Comments:* Theorem 1 provides the sufficient condition for a $\Upsilon_{v_1, v_2}$ CGR to generate an MDS code (and its dual). When those conditions are satisfied, Theorem 2 provides the concrete procedure to construct the $t$-erasure correcting MDS code. The proof of Theorem 2 is rather involved. Due to the space limitation, only a sketch is provided. A key to prove that the array code constructed from Algorithm 1,2,3 is indeed MDS is by interpreting the complete-graph-of-ring as a "ring-of-complete-graph", namely, a set of complete graphs connected by ring edges. The viewpoint allows us to slide the CGR graphs to slices of disjoint subgraphs, each of which embeds a $B_{2n+1}$-code except for two slices [1]. Using the fact that $B_{2n+1}$-code is MDS, we can easily prove, except for the two special slices, that any missing $t$ columns (the components in the missing columns that belong to the respective subgraph) can be recovered from the remainder. ! The two special slices do not individually possess the $t$-erasure correcting ability, but combined together, they can be proven to successfully handle $t$ column erasures. Gathering all these pieces together leads to conclusion of MDS. The detailed proof is omitted due to space limitation, but the discussion in Section IV (which relates CGR codes to $B_{2n+1}$ codes) shed insight into the proof.

## III. DUAL CODES

We have thus far provide a concrete method to arrange all the edges and vertices in CGR($K_{v_1}, C_{v_2}$) in an array of $v_1 v_2 / 2 \times v_2$. Using the computer search, more MDS arrangements can be found, Table III provides some examples. The first offset vector for each configuration is obtained using Algorithm 3, and the rest offset vectors are from computer search. By interpreting vertices as information bits and edges as parity bits, we obtained a rich class of MDS codes capable of correcting $(v_1 + 1)$ erasures for $v_1 = 2, 4, 6....$. Using the same array layout, but reversing the roles of vertices and edges, we get their respective dual codes, which are always 2-erasure-correcting.

**Example 4**: The $5 \times 5$ array code in Example 2 is a $(5, 2)$ 3-erasure-correcting MDS code constructed from CGR($K_2, C_5$) in Fig. 2. The same array arrangement can be mapped to a dual MDS code with 2-erasure-correcting capability, by reversing the roles of edges and vertices (i.e. letting edges represent data bits and vertices represent parity bits). To ease the representation, we re-label the edges using alphabets $a, b, ...o$ as in Fig. 2, and interpret

| $K_{v_1}$ | $C_{v_2}$ | array size | offset vector | code $(n,k)$ |
|---|---|---|---|---|
| $K_2$ | $C_5$ | $5 \times 5$ | 0,1,2,2,4 | (5,2) |
| $K_4$ | $C_7$ | $14 \times 7$ | 0,1,2,3,4,4,4,4,3,6,2,0,6,1 | (7,2) |
| $K_4$ | $C_7$ | $14 \times 7$ | 0,1,2,3,4,4,4,4,6,1,2,3,0,6 | (7,2) |
| $K_4$ | $C_7$ | $14 \times 7$ | 0,1,2,3,4,4,4,4,6,3,1,0,2,6 | (7,2) |
| $K_4$ | $C_7$ | $14 \times 7$ | 0,1,2,3,5,5,5,5,2,3,4,4,0,1 | (7,2) |
| $K_6$ | $C_9$ | $27 \times 9$ | 0,1,2,3,4,5,6,6,6,6,6,6,2,3,1,5,8,4,8,0,3,5,8,0,2,4,1 | (9,2) |
| $K_6$ | $C_9$ | $27 \times 9$ | 0,1,2,3,4,5,6,6,6,6,6,6,2,3,1,5,8,4,8,3,0,5,8,1,0,4,2 | (9,2) |
| $K_6$ | $C_9$ | $27 \times 9$ | 0,1,2,3,4,5,6,6,6,6,6,6,2,3,4,8,1,5,8,3,4,1,0,8,5,0,2 | (9,2) |
| $K_8$ | $C_{11}$ | $44 \times 11$ | 0,1,2,3,4,5,6,7,8,8,8,8,8,8,8,8,2,3,4,1,6,7,10,10,5,3,7,0,4,6,7,1,4,5,10,2,1,0,0,5,6,10,3,2 | (11,2) |

TABLE I
PARAMETERS FOR MDS-CGR ARRAY CODES

the vertices of degree 3 as parities on 3 information bits. The $(5,3)$ dual code takes the following form:

| $a \oplus l \oplus e$ | $a \oplus m \oplus b$ | $b \oplus n \oplus c$ | $c \oplus o \oplus d$ | $d \oplus k \oplus e$ |
|---|---|---|---|---|
| $m \oplus g \oplus h$ | $h \oplus n \oplus i$ | $o \oplus i \oplus j$ | $j \oplus k \oplus f$ | $g \oplus l \oplus f$ |
| c | d | e | a | b |
| i | j | f | g | h |
| k | l | m | n | o |

## IV. CONNECTION TO B-CODES

We now reveal an interesting relation between the proposed CGR codes and B-codes $B_{2n+1}$: namely, stripping off certain rows and columns from the CGR array results in a $B_{2n+1}$ code. On one hand, since CGR graphs are nested graphs with a complete graph as the base graph, and since B-codes can be constructed from complete graphs, it should not be surprising that CGR codes subsumes B-codes as contracted codes. On the other hand, in addition to the significantly more complex structure of CGR codes, another notable difference is that CGR codes are by nature cyclically symmetric, whereas $B_{2n+1}$ codes have an asymmetric structure as shown in Fig. 5. This difference makes the connection between CGR codes and $B_{2n+1}$ codes all the more interesting, and we will discuss this through the following example.

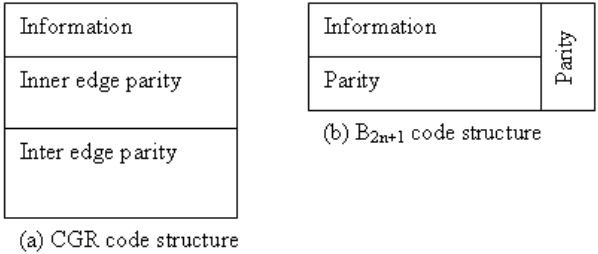

Fig. 5. (a) Structure of CGR code. (b) Structure of $B_{2n+1}$ code

**Example 5:** Consider a $(7,2)$ MDS array code constructed from CGR$(K_4, C_7)$ using the algorithms discussed before and the offset vector $(0,1,2,3,4,4,4,4,2,3,6,6,0,1)$ constructed in Example 3. The code is listed in Fig. II (a).

To contract a complete-graph-of-ring back to a complete graph, consider only the first vertex of each ring and the edges connecting these vertices only. Puncturing all other vertices and edges leads to Fig. II (b).

Next, vertically and horizontally compact the array and reorder the columns, we get a $(5,2)$ $B_5$ code from the original $(7,2)$ CGR code:

| 0 | 7 | 14 | 21 | $0 \oplus 21$ |
|---|---|---|---|---|
| $7 \oplus 21$ | $14 \oplus 21$ | $0 \oplus 7$ | $0 \oplus 14$ | $7 \oplus 14$ |

We note that puncturing and shortening the CGR code to a B-code is straight-forward, but the reverse operation of expanding a B-code to a CGR code is nontrivial. Nevertheless, the above relation is inspiring as it sheds insight into the possibility of constructing new and more sophisticated codes from existing codes.

## V. PERFORMANCE ANALYSIS

In this section, the update and decoding complexity are concisely considered. From our CGR code construction above, note that each column in the array (a storage unit) consists of both systematic bits and parity bits.

**Definition 4:** The **update complexity** is defined as the number of parity updates required while a single information bit is changed or updated, averaged over all the information (systematic) bits [9].

**Definition 5:** The **decoding complexity** is defined as the number of bit operations (e.g. XOR, AND, shift) required in order to recover the erased symbols (columns) from the survivors, averaged over all the formation symbols.

Recall that the proposed CGR codes based on $K_{v_1}$ and $C_{v_2}$ where $v_2 = v_1 + 3$. For updating a single information bit, since every information bits involves to $v_1 + 1$ parity bits, it will give rise to the update of $(v_1 + 1)$ parity bits. Hence, averaged over $v_1 v_2$ information bits, the update complexity will be $(v_2 - 2)/(v_2(v_2 - 3))$. The update complexity for different code configurations is listed in Table I. Since the code is one with parameters $(n,k) = (v_2, 2)$, the update complexity decreases linearly with the "code-length" $v_2$ and goes to zero asymptotically.

To compute the decoding complexity, we can consider that all the $v_1 v_2 / 2$ bits in an arbitrary missing column takes one XOR operation per bit, or, one XOR operation for the entire symbol. In the worst case, the code has a payload of two systematic symbols (or $v_1 v_2$ systematic bits), so the decoding complexity is $\frac{1}{2}$ per erased symbol, irrespective of the code lengths.

(a) The array layout of a CGR code

| **0** | 1 | 2 | 3 | 4 | 5 | 6 |
|---|---|---|---|---|---|---|
| 8 | 9 | 10 | 11 | 12 | 13 | **7** |
| 16 | 17 | 18 | 19 | 20 | **14** | 15 |
| 24 | 25 | 26 | 27 | **21** | 22 | 23 |
| $4 \oplus 5$ | $5 \oplus 6$ | $6 \oplus 0$ | $\mathbf{0 \oplus 1}$ | $1 \oplus 2$ | $2 \oplus 3$ | $3 \oplus 4$ |
| $11 \oplus 12$ | $12 \oplus 13$ | $13 \oplus 7$ | $\mathbf{7 \oplus 8}$ | $8 \oplus 9$ | $9 \oplus 10$ | $10 \oplus 11$ |
| $18 \oplus 19$ | $19 \oplus 20$ | $20 \oplus 14$ | $\mathbf{14 \oplus 15}$ | $15 \oplus 16$ | $16 \oplus 17$ | $17 \oplus 18$ |
| $25 \oplus 26$ | $26 \oplus 27$ | $27 \oplus 21$ | $\mathbf{21 \oplus 22}$ | $22 \oplus 23$ | $23 \oplus 24$ | $24 \oplus 25$ |
| $2 \oplus 9$ | $3 \oplus 10$ | $4 \oplus 11$ | $5 \oplus 12$ | $6 \oplus 13$ | $\mathbf{0 \oplus 7}$ | $1 \oplus 8$ |
| $3 \oplus 17$ | $4 \oplus 18$ | $5 \oplus 19$ | $6 \oplus 20$ | $\mathbf{0 \oplus 14}$ | $1 \oplus 15$ | $2 \oplus 16$ |
| $6 \oplus 27$ | $\mathbf{0 \oplus 21}$ | $1 \oplus 22$ | $2 \oplus 23$ | $3 \oplus 24$ | $4 \oplus 25$ | $5 \oplus 26$ |
| $13 \oplus 20$ | $\mathbf{7 \oplus 14}$ | $8 \oplus 15$ | $9 \oplus 16$ | $10 \oplus 17$ | $11 \oplus 18$ | $12 \oplus 19$ |
| $\mathbf{7 \oplus 21}$ | $8 \oplus 22$ | $9 \oplus 23$ | $10 \oplus 24$ | $11 \oplus 25$ | $12 \oplus 26$ | $13 \oplus 27$ |
| $15 \oplus 22$ | $16 \oplus 23$ | $17 \oplus 24$ | $18 \oplus 25$ | $19 \oplus 26$ | $20 \oplus 27$ | $\mathbf{14 \oplus 21}$ |

(b) The array layout of a B-code (reduced from the CGR code)

| 0 | - | - | - | - | - | - |
|---|---|---|---|---|---|---|
| - | - | - | - | - | - | 7 |
| - | - | - | - | - | 14 | - |
| - | - | - | - | 21 | - | - |
| - | - | - | - | - | - | - |
| - | - | - | - | - | - | - |
| - | - | - | - | - | - | - |
| - | - | - | - | - | - | - |
| - | - | - | - | - | $0 \oplus 7$ | - |
| - | - | - | - | $0 \oplus 14$ | - | - |
| - | $0 \oplus 21$ | - | - | - | - | - |
| - | $7 \oplus 14$ | - | - | - | - | - |
| $7 \oplus 21$ | - | - | - | - | - | - |
| - | - | - | - | - | - | $14 \oplus 21$ |

TABLE II

RELATION BETWEEN CGR CODES AND B-CODES

| $K_{v_1}$ | $C_{v_2}$ | code | update complexity | decode complexity |
|---|---|---|---|---|
| $K_2$ | $C_5$ | (5,2) | $3/10 = 30\%$ | $1/2 = 50\%$ |
| $K_4$ | $C_7$ | (7,2) | $5/28 = 18\%$ | $1/2 = 50\%$ |
| $K_6$ | $C_9$ | (9,2) | $7/54 = 13\%$ | $1/2 = 50\%$ |
| $K_8$ | $C_{11}$ | (11,2) | $9/88 = 10\%$ | $1/2 = 50\%$ |
| $K_{10}$ | $C_{13}$ | (13,2) | $11/130 = 8\%$ | $1/2 = 50\%$ |

TABLE III

UPDATE COMPLEXITY AND DECODING COMPLEXITY

## VI. CONCLUSION

We have presented a concrete method to construct a rich class of $(2k+3, 2)$ array codes and their $(2k+3, 2k+1)$ dual codes, for $k = 1, 2, \ldots$. These codes are capable of correcting up to $3, 5, 7, 9, \ldots$ erasures, and their dual codes, which follow straight-forward from the same configuration, can correct up to 2 erasures. These codes grow out of a new class of graphs known as the complete-graph-of-rings. Instead of computer search, we demonstrate a deterministic methods that can leverage from perfect one-factorization. We further show that the $B_{2n+1}$ codes can be obtained by trimming down the CGR codes.